# A case for resonant x-ray Bragg diffraction by a collinear antiferromagnet $Li_2Ni_3P_4O_{14}$

S. W. Lovesey

*ISIS Facility, STFC, Didcot, Oxfordshire OX11 0QX, UK*
*Diamond Light Source, Harwell Science and Innovation Campus, Didcot, Oxfordshire OX11 0DE, UK*
*Department of Physics, Oxford University, Oxford OX1 3PU, UK*

**Abstract** Magnetic axial and polar (Dirac) nickel multipoles contribute to resonant x-ray Bragg amplitudes in a symmetry informed analysis of monoclinic $Li_2Ni_3P_4O_{14}$ presented for future diffraction experiments. Magnetic long-range order below a temperature ≈ 14.5 K can be viewed as a two-dimensional trimerized antiferromagnet with Ni ions in two Wyckoff positions in the centrosymmetric ($\bar{1}$) magnetic space group $P2_1/c$. It permits the coupling to circular polarization in the primary x-ray beam, unlike the corresponding diffraction by an antiferromagnet characterized by anti-inversion ($\bar{1}'$) and a linear magnetoelectric effect, e.g., historically significant chromium sesquioxide ($Cr_2O_3$) and $Cu_2(MoO_4)(SeO_3)$ (Lovesey & van der Laan, 2024). The space group is inferred from neutron Bragg diffraction patterns, without an allowance for permitted Dirac dipoles (anapoles) and quadrupoles (Chikara *et al*., 2025).

## 1. Introduction

Electronic correlations in conjunction with spin-orbit coupling in solids create valuable and enigmatic effects (Pourovskii *et al*., 2025). These include the magnetoelectric effect, collinear antiferromagnetic order, and so-called hidden orders in which a state of matter exists without an easily detectable order parameter. Scattering techniques often have much to offer in gathering incisive information on properties of solids not available with other experimental methods. For example, Néel (1932) proposed antiferromagnetism and J. Samuel Smart (1949) recognized that magnetic neutron Bragg diffraction could provide the concrete evidence. To this end, diffraction patterns were collected from powdered MnO that presents magnetic order below a temperature ≈ 122 K. A difference in patterns from samples at room temperature and a temperature ≈ 80 K revealed strong magnetic Bragg spots at positions not allowed on the basis of the chemical face-centred cubic unit cell of MnO.

A meaningful analysis of a Bragg diffraction pattern requires the appreciation of symmetry properties of the radiation-matter interaction and the illuminated sample (Colins *et al*., 2007; Lovesey & Balcar, 2013; Winkler & Zülicke, 2025). For the sample, an extended Neumann's principle (Neumann, 1885; Cracknell, 1975) imposes space inversion symmetry and time inversion symmetry for a resonant ion. Translations absent in the magnetic crystal class appear in the magnetic space group [Bilbao Crystallographic server, http://www.cryst.ehu.es, Belov-Neronova-Smirnova (BNS) setting of magnetic space groups]. Discrete symmetries of the radiation-matter interaction are space inversion symmetry, time inversion symmetry and photon polarization. Crystals of berlinite (K absorption edge of Al), tellurium ($L_1$-edge) and quartz (K-edge of Si) belong to space groups $P3_121$ (No. 152 right-handed screw) or $P3_221$ (No. 154 left-handed screw), and an absolute chirality is verified by Bragg diffraction of circularly polarized x-rays tuned in energy to the specified absorption edges (Tanaka *et al*., 2010). The most probable absorption event is the parity-even (axial) electric dipole-electric dipole (E1-E1) with atomic transitions 1s → 3p for Al and Si, and 2s → 5p for Te. The parity-odd (polar) electric dipole - electric quadrupole (E1-E2) event plays a minor role. Parity-even E1-E1 and E2-E2 absorption events include non-magnetic (time inversion symmetry even) and magnetic (time inversion symmetry odd) processes. The latter can reveal the long-range magnetic order of familiar axial dipole moments. The corresponding dipoles in a parity-odd E1-E2 absorption event are less familiar anapoles that possess discrete symmetries akin to that of a Dirac monopole (Milton, 2006).

A magnetic ground state for $Li_2Ni_3P_4O_{14}$ depicted in Fig. 1 has been proposed on the basis of magnetic susceptibility and specific heat measurements, and neutron-powder diffraction and neutron polarization analysis (Chikara *et al*., 2025). Magnetic long-range order below a temperature ≈ 14.5 K can be viewed as a two-dimensional trimerized antiferromagnet involving two independent Ni ions. The monoclinic magnetic space group assigned by the authors possesses a propagation vector = (0, 0, 0), commensurate collinear antiferromagnetic order and a ferromagnetic component parallel to the unique crystal axis. The permitted coupling of the magnetic order to photon circular polarization in the primary beam is one delineated characteristic of symmetry informed amplitudes for resonant x-ray Bragg diffraction. Furthermore, Wyckoff positions in Fig. 1 differ with respect to spatial inversion symmetry. Polar magnetic multipoles are permitted at the acentric position, including anapoles visible in resonant x-ray and neutron Bragg diffraction patterns (Fernández-Rodríguez *et al*., 2010; Lovesey *et al*., 2019).

Resonant x-ray Bragg diffraction by magnetic $V_2O_3$ illustrates the puissance of the technique (Paolasini *et al*., 1999; Paolasini *et al*., 2001; Lovesey & Knight, 2000). Bragg spots forbidden in the diffraction pattern of the monoclinic structure are due to orbital magnetism and not orbital-order as originally proposed. The experiment exploited a relatively weak E2 absorption event at the vanadium K-edge. Turning to a second example and a high-$T_c$ material, emergence of the pseudo-gap phase with time-reversal violation in underdoped $HgBa_2CuO_{4+\delta}$ (Hg1201) diminishes Cu site symmetry. Specifically, the Cu centre of inversion symmetry is lost, and Dirac multipoles herald the onset of the enigmatic phase (Lovesey & Kalyavin, 2015). Analysis of magnetic neutron diffraction patterns for Hg1201 yield the orthorhombic magnetic symmetry of the pseudo-gap phase (Bourges *et al*., 2021; Croft *et al*., 2017; Bourges *et al*., 2018; Fechner *et al*., 2016).

## 2. Magnetic structure

The magnetic structure $P2_1/c$ (No.14.75, Bilbao) adopted for $Li_2Ni_3P_4O_{14}$ is depicted in Fig. 1 with nickel ions ($Ni^{2+}$, $3d^8$) in Wyckoff general positions 4e and special positions 2c

(Chikara *et al*., 2025). Sites 4e are devoid of any symmetry and 2c are centrosymmetric. Nickel ions in positions 2c are permitted axial (parity even) multipoles alone, while positions 4e are permitted axial and Dirac multipoles. Reflection conditions include (0, 2$n$, 0) and (0, 0, 2$n$), and the monoclinic cell possesses unique axis b (Bilbao). Magnetic multipoles are referred to orthogonal axes labelled (ξ, η, ζ) derived in standard form, namely, ξ ∝ **a**\*, η ∝ **b**, ζ ∝ **c**, with reciprocal lattice vectors **a**\* ∝ $a$(1, 0, 0), **b**\* ∝ $b$(0, 1, 0), **c**\* ∝ $c$(− cos(β$_o$), 0, sin(β$_o$)) and an obtuse angle β$_o$ ≈ 110.31° (Chikara *et al*., 2025).

In our adopted description of electronic degrees of freedom, Ni ions are assigned spherical multipoles ⟨$O^K_Q$⟩ of integer rank K with projections Q in the interval −K ≤ Q ≤ K (Lovesey *et al*., 2005; Lovesey & Balcar, 2013). Angular brackets denote the time-average, or expectation value, of the enclosed spherical tensor operator. Cartesian and spherical components Q = 0, ±1 of a vector **n** = (ξ, η, ζ) are related by ξ = ($n_{-1}$ − $n_{+1}$)/√2, η = $i$($n_{-1}$ + $n_{+1}$)/√2, ζ = $n_0$. A complex conjugate of a multipole is defined as ⟨$O^K_Q$⟩\* = $(-1)^Q$ ⟨$O^K_{-Q}$⟩, meaning the diagonal multipole ⟨$O^K_0$⟩ is purely real. The phase convention for real and imaginary parts labelled by single and double primes is ⟨$O^K_Q$⟩ = [⟨$O^K_Q$⟩′ + $i$⟨$O^K_Q$⟩″]. Whereupon Cartesian dipoles are ⟨$O^1_ξ$⟩ = −√2 ⟨$O^1_{+1}$⟩′ and ⟨$O^1_η$⟩ = −√2 ⟨$O^1_{+1}$⟩″.

A structure factor (Lovesey *et al*., 2005),

$$\Psi^K_Q = [\exp(i\mathbf{\kappa} \cdot \mathbf{d}) \langle O^K_Q \rangle_\mathbf{d}], \quad (1)$$

delineates the Bragg diffraction pattern for a reflection vector **κ** defined by integer Miller indices ($h$, $k$, $l$). The sum is over positions **d** used by Ni ions in P2$_1$/c (No. 14.75, magnetic crystal class 2/m). Reflection conditions are derived from $\Psi^K_Q$ by considering its value for even K and Q = 0, and a parity-even signature σ$_π$ = +1 (the structure factor for nuclear scattering). Bulk magnetic properties are defined by $\Psi^K_Q$ evaluated for K = 1 (dipole), σ$_π$ = +1 and $h$ = $k$ = $l$ = 0.

In more detail, Eq. (1) requires information about the relevant Wyckoff positions found in the Bilbao table MWYCKPOS for the magnetic symmetry of interest. Wyckoff positions are related by operations listed in the table MGENPOS (Bilbao). Taken together, the two tables provide all information required to evaluate Eq. (1) and, thereafter, all x-ray diffraction amplitudes.

Henceforth, axial (σ$_π$ = +1) and polar (σ$_π$ = −1) multipoles are denoted by ⟨$T^K_Q$⟩ and ⟨$G^K_Q$⟩, respectively. The time signature σ$_θ$ of ⟨$T^K_Q$⟩ is σ$_θ$ = $(-1)^K$, i.e., even K axial multipoles are non-magnetic. Dirac multipoles ⟨$G^K_Q$⟩ are magnetic with σ$_θ$ = $(-1)$ for all K. We find,

$$\Psi^K_Q(4e) = \langle O^K_Q \rangle [\alpha\beta\gamma + \sigma_\pi (\alpha\beta\gamma)^*] + (-1)^{K+Q} (-1)^{k+l} \langle O^K_{-Q} \rangle [(\alpha\gamma)^*\beta + \sigma_\pi (\alpha\gamma)\beta^*], \quad (2)$$

Spatial phase factors in Eq. (2) are α = exp($i2\pi h$x), β = exp($i2\pi k$y) and γ = exp($i2\pi l$z). Estimates of the general coordinates (x, y, z) for Ni ions are given by Chikara *et al*. (2025). Wyckoff positions 2c are centres of inversion symmetry ($\bar{1}$, $\sigma_\pi$ = +1) and no more. The result,

$$\Psi^K_Q(2c) = \langle T^K_Q \rangle + (-1)^{k+l} \langle T^K_{-Q} \rangle, \qquad (3)$$

complies with a reflection condition $(k + l) = 2n$.

## 3. X-ray diffraction

Tuning the energy of the x-rays to an atomic resonance has two obvious benefits in diffraction experiments (Paolasini, 2014). In the first place, there is a welcome enhancement of Bragg spot intensities and, secondly, spots are element specific. States of x-ray polarization, Bragg angle θ, and the plane of scattering are shown in Fig. 2. A conventional labelling of linear photon polarization states places **σ** = (0, 0, 1) and **π** = (cos(θ), sin(θ), 0) perpendicular and parallel to the plane of scattering, respectively. Secondary states **σ′** = **σ** and **π′** = (cos(θ), − sin(θ), 0).

The x-ray scattering length in the unrotated channel of polarization σ → σ′, say, is modelled by (σ′σ)/D(E). In this instance, the resonant denominator is replaced by a sharp oscillator D(E) = {[E − Δ + $i\Gamma/2$]/Δ} with the x-ray energy E in the near vicinity of an atomic resonance Δ of total width Γ, namely, E ≈ Δ and Γ << Δ. The cited energy-integrated scattering amplitude (σ′σ), one of four amplitudes, is studied using standard tools and methods from atomic physics and crystallography. In the first place, a vast spectrum of virtual intermediate states makes the x-ray scattering length extremely complicated. It can be truncated following closely steps in celebrated studies by Judd and Ofelt of optical absorption intensities of rare-earth ions (Judd, 1962; Ofelt, 1962; Hehlen *et al*., 2013). An intermediate level of truncation used here reproduces sum rules for axial dichroic signals created by E1-E1 or E2-E2 absorption events (Johnson & Lovesey, 2024). The attendant calculation presented in Lovesey & Balcar (1997) and Section 5.2 in Lovesey *et al*. (2005) is lengthy and demanding. Experimental results for Dirac multipoles in $V_2O_3$ (Fernández-Rodríguez *et al*., 2010) and CuO (Scagnoli *et al*., 2011; Lovesey & Balcar, 2013) have been published together with successful interpretations. The study of $V_2O_3$ is noted for a full use of linear photon polarization analysis to disentangle multipoles in diffraction amplitudes.

Here, we implement universal expressions for scattering amplitudes and abbreviate notation using (σ′σ) ≡ $F_{\sigma'\sigma}$, etc., for E1-E1 amplitudes listed by Scagnoli & Lovesey (2009), Appendix C. Likewise, E1-E2 amplitudes by Scagnoli & Lovesey (2009), Appendix D, and E1-M1 where M1 is the magnetic dipole moment (Lovesey & Balcar, a 2010; Lovesey & Balcar, b 2010).

Laue conditions for magnetic reflections (0, 1, 0) and (0, 0, 1) are satisfied at the nickel $L_3$ and $L_2$ absorption edges at E ≈ 0.861 keV and E ≈ 0.878 keV, respectively. The relevant lattice parameters for $Li_2Ni_3P_4O_{14}$ are $b \approx 7.749$ Å and $c \approx 9.337$ Å (Chikara *et al*., 2025), e.g., $\sin(\theta) = [\lambda/\{2c \sin(\beta_o)\}]$ for (0, 0, 1) with a photon wavelength $\lambda \approx (12.4/E)$ Å. An E1 event $2p \rightarrow 3d$ is much stronger than E1 ($1s \rightarrow 4p$) and E2 ($1s \rightarrow 3d$) events at the nickel K edge at $E \approx 8.339$ keV (Paolasini, 2014).

## 4. Reflections (0, 2*n* + 1, 0)

The crystal axis c is normal to the plane of scattering depicted in Fig. (2). There is no E1-E1 diffraction in the unrotated σ-channel i.e., (σ′σ) = 0. Two remaining (0, 2*n* + 1, 0) diffraction amplitudes, apart from common factors {4 cos(2π*k*y)} with y ≈ 0.1314 (Chikara *et al*., 2025) for positions 4e or 2 for positions 2c, are,

$$(\pi'\pi) = i \sin(2\theta) [\cos(\psi) \langle T^1\zeta\rangle - \sin(\psi) \langle T^1\xi\rangle],$$

(4)

$$(\pi'\sigma) = \cos(\theta) [-i\{\sin(\psi) \langle T^1\zeta\rangle + \cos(\psi) \langle T^1\xi\rangle\} + \cos(\psi) \langle T^2_{+1}\rangle'' + \sin(\psi) \langle T^2_{+2}\rangle''].$$

The rotated channel amplitude (π′σ) demonstrates that contributions from axial magnetic dipoles and non-magnetic Templeton-Templeton (T & T) scattering are shifted in phase by 90º (Templeton & Templeton, 1985; Templeton & Templeton, 1986; Ovchinnikova *et al*., 2025). Scattered intensity picked out by circular polarization in the primary photon beam = $P_2\Upsilon$ with a chiral signature (Tanaka *et al*., 2010),

$$\Upsilon = \{(\sigma'\pi)^*(\sigma'\sigma) + (\pi'\pi)^*(\pi'\sigma)\}'',$$

(5)

and the Stokes parameter $P_2$ (a purely real pseudoscalar) measures helicity in the primary x-ray beam. Since intensity is a true scalar, $\Upsilon$ and $P_2$ must possess identical discrete symmetries, specifically, both scalars are time-even and parity-odd. The signature is extracted from observed intensities by subtraction of intensities measured with opposite handed primary x-rays, namely, ± $P_2$. Intensity of a Bragg spot in the rotated channel of polarization is proportional to |(π′σ)|², and likewise for unrotated channels of polarization. The corresponding E1-E1 chiral signature is,

$$\Upsilon(4e) = \{(\pi'\pi)^*(\pi'\sigma)\}'' = \cos^2(2\pi k y) \cos(\theta) \sin(2\theta) [\sin(\psi) \langle T^1\xi\rangle - \cos(\psi) \langle T^1\zeta\rangle]$$

$$\times [\cos(\psi) \langle T^2_{+1}\rangle'' + \sin(\psi) \langle T^2_{+2}\rangle''].$$

(6)

Notably, the chiral signature is a product of dipoles and T & T scattering, and a function of sin(2ψ) and cos(2ψ).

An E1-E2 absorption event reveals Dirac multipoles $\langle G^K\rangle$ with K = 1, 2, 3 hosted by Wyckoff positions 4e. We consider the unrotated amplitude (σ′σ) in light of the absence of diffraction in this channel for an E1-E1 event. The Dirac amplitude is purely imaginary and proportional to $\{\sin(2\pi ky) \cos(\theta)\}$. Working to the level of the diagonal octupole,

$$(\sigma'\sigma) \approx i \sin(2\pi ky) \cos(\theta) [\sin(\psi) \langle G^1_\zeta\rangle + \cos(\psi) \langle G^1_\xi\rangle$$

$$+ (\sqrt{10}/3) \{\cos(\psi) \langle G^2_{+1}\rangle'' + \sin(\psi) \langle G^2_{+2}\rangle''\} + \sqrt{(2/3)} \sin(\psi) (5 \cos^2(\psi) - 1) \langle G^3_0\rangle + \ldots]. \quad (7)$$

The three omitted octupoles are $\langle G^3_Q\rangle'$ with Q = 1 - 3. Anapoles (K = 1) in $(\sigma'\sigma)$ are parallel to **a*** and **c**. The four amplitudes have identical phases, namely, purely imaginary, and the E1-E2 chiral signature is zero.

**5. Reflections** $(0, 0, 2n + 1)$

Diffraction amplitudes for space group forbidden reflections $(0, 0, 2n + 1)$ depend on the monoclinic obtuse angle, and we use $d = \cos(\beta_o)$ and $e = \sin(\beta_o)$ for an E1-E1 absorption event. It is convenient to employ multipoles,

$$A_1 = [d \langle T^1_\zeta\rangle + e \langle T^1_\xi\rangle], \quad B_1 = [e \langle T^1_\zeta\rangle - d \langle T^1_\xi\rangle],$$

$$A_2 = [d \langle T^2_{+1}\rangle'' - e \langle T^2_{+2}\rangle''], \quad B_2 = [e \langle T^2_{+1}\rangle'' + d \langle T^2_{+2}\rangle'']. \quad (8)$$

Axial dipoles in $A_1$ and $B_1$ are parallel to **a*** and **c**, and T & T scattering create $A_2$ and $B_2$.

The cited amplitudes possess common factors $\{4 \cos(2\pi lz)\}$ with z ≈ 0.9789 (Chikara *et al*., 2025) or 2 for Wyckoff positions 4e and 2c, respectively. The unique axis **b** is parallel to the axis y in Fig. 2 at the start of an azimuthal angle scan $\psi = 0$. E1-E1 amplitudes are,

$$(\sigma'\sigma) = \sin(2\psi) A_2,$$

$$(\pi'\pi) = (i/\sqrt{2}) \sin(2\theta) \cos(\psi) A_1 + \sin^2(\theta) \sin(2\psi) A_2, \quad (9)$$

$$(\pi'\sigma) = (i/\sqrt{2}) [-\cos(\theta) \sin(\psi) A_1 + \sin(\theta) B_1] + \sin(\theta) \cos(2\psi) A_2 - \cos(\theta) \sin(\psi) B_2.$$

The unrotated amplitude $(\sigma'\sigma)$ is non-magnetic T & T scattering, and it is independent of the Bragg angle θ. Multipoles $A_1$ and $A_2$ in the amplitudes $(\pi'\sigma)$ and $(\sigma'\pi)$ take opposite signs. The chiral signature is caused by an interference between axial dipoles and T & T scattering,

$$\Upsilon(4e) = \cos^2(2\pi lz) \cos(\theta) [(B_1 A_2 - A_1 B_2) \sin(2\theta) \sin(2\psi)$$

$$+ 2 A_1 A_2 \{(1 + \sin^2(\theta)) \sin(\psi) \sin(2\psi) + 2 \sin^2(\theta) \cos(\psi) \cos(2\psi)\}]. \quad (10)$$

The expression $\Upsilon(4e)$ is a sum of odd and even functions of the azimuthal angle.

An E1-E2 absorption event reveals the anapole parallel to the unique axis **b** in the unrotated amplitude $(\sigma'\sigma)$,

$$(\sigma'\sigma) \approx i \sin(2\pi lz) \cos(\theta) \cos(\psi) [-\langle G^1_\eta \rangle$$

$$+ (1/3) \sqrt{(5/2)} \{\sin(2\beta_o) (\langle G^2_0 \rangle - \langle G^2_{+2} \rangle') + 2 \cos(2\beta_o) \langle G^2_{+2} \rangle'\} + ]. \quad (11)$$

Octupoles omitted in Eq. (11) are $\langle G^3_Q \rangle''$ with Q = 1 - 3. In line with reflections $(0, 2n+1, 0)$, the four amplitudes have identical phases and the E1-E2 chiral signature is zero.

## 6. Summary and discussion

We have studied Bragg diffraction patterns for the low temperature phase of $Li_2Ni_3P_4O_{14}$. On the basis of an extensive set of investigations, Chikara *et al*. (2025) conclude that it is a monoclinic, collinear antiferromagnet with nickel ions using two Wyckoff positions in the magnetic space group $P2_1/c$ (No. 14.75) depicted in Fig. 1. Symmetry informed calculations of resonant x-ray Bragg diffraction patterns reported in the main text include Dirac multipoles (polar and magnetic) permitted at the position that is asymmetric. Anapoles (Dirac dipoles) engaged in diffraction are parallel to the unique axis **b**, and **a*** and **c**, respectively, for two classes of space group forbidden reflections. The magnetic crystal class (2/m) permits the coupling of the magnetic structure to circular polarization (helicity) in the primary beam of x-rays using a parity-even electric-dipole electric-dipole (E1-E1) absorption event. In our study, the coupling is quantified by a chiral signature proportional to the change in x-ray helicity on scattering (Tanaka *et al*., 2010). The chiral signature is zero for Dirac multipoles exposed by a parity-odd electric-dipole electric-quadrupole (E1-E2) absorption event. Calculated diffraction patterns are enriched by simulating azimuthal angle scans in which the crystal is rotated about the reflection vector.

Chikara *et al*. (2025) did not allow Dirac multipoles in the analysis of their magnetic Bragg diffraction patterns gathered with neutron scattering (Lovesey *et al*., 2019). The technique measures the spatial distribution of magnetization (Brown, 1993). Their published measurements extend to a small d-spacing (wavevector $\kappa \approx 1.36$ Å$^{-1}$) that spans a maximum in the radial integral that accompanies an anapole and a Dirac quadrupole (Lovesey, 2015; Lovesey & van der Laan, 2024).

**Acknowledgements** Correspondence with Dr A. K. Bera. Dr D. D. Khalyavin commented on the study in its making and prepared Fig. 1.

## References

Bourges, P., Bounoua, D. & Sidis, Y. (2021). Comptes Rendus Physique **22**, 1.

Bourges, P., Sidis, Y. & Mangin-Thro, L. (2018). Phys. Rev. B **98**, 016501.

Brown, P. J. (1993). Int. J. Mod. Phys. B **7**, 3029.

Chikara, K. S., Bera, A. K., Kumar, A., Yusuf, S. M., Orlandi, F. & Balz, C. (2025). Phys. Rev. B **112**, 014438.

Colins, S. P., Lovesey, S. W. & Balcar, E. (2007). J. Phys.: Condens. Matter **19**, 213201-09.

Cracknell, A. P. (1975). *Magnetism in Crystalline Materials*, Pergamon Press, Oxford.

Croft, T. P., Blackburn, E., Kulda, J., Liang, R., Bonn, D. A., Hardy, W. N. & Hayden, S. M. (2017). Phys. Rev. B **96**, 214504.

Fechner, M., Fierz, M. J. A., Thöle, F., Staub, U. & Spaldin, N. A. (2016). Phys. Rev. B **93**, 174419.

Fernández-Rodríguez, J., Scagnoli, V., Mazzoli, C., Fabrizi, F., Lovesey, S. W., Blanco, J. A., Sivia, D. S., Knight, K. S., de Bergevin, F. & Paolasini, L. (2010). Phys. Rev. B **81**, 085107.

Hehlen, M. P., Brik, M. G. & Krämer, K. W. (2013). J. Luminescence **136**, 221.

Johnson, R. D. & Lovesey, S. W. (2024). Phys. Rev. B **110**, 104405.

Judd, B. R. (1962). Phys. Rev. **127**, 750.

Lovesey S. W. & Balcar E. (1997). J. Phys.: Condens. Matter **9**, 4237.

Lovesey, S. W. & Balcar, E. a (2010). J. Phys. Soc. Jpn. **79**, 074707.

Lovesey, S. W. & Balcar, E. b (2010). J. Phys. Soc. Jpn. **79**, 104702.

Lovesey, S. W. & Balcar, E. (2013). J. Phys. Soc. Jpn. **82**, 021008.

Lovesey, S. W., Balcar, E., Knight, K. S. & Fernández-Rodríguez, J. (2005). Phys. Rep. **411**, 233.

Lovesey, S. W. & van der Laan, G. (2024). Phys. Rev **B**,110.174442.

Lovesey, S. W. (2015). Phys. Scr. **90**, 108011.

Lovesey, S. W., Chatterji, T., Stunault, A., Khalyavin, D. D. & van der Laan, G. (2019). Phys. Rev. Lett. **122**, 047203.

Lovesey S. W. & Knight, K. S. (2000). J. Phys.: Condens. Matter **12**, L367.

Lovesey, S. W. & Khalyavin, D. D. (2015). J. Phys.: Condens. Matter **27**, 495601.

Milton, K. A. (2006). Rep. Prog. Phys. **69**, 1637.

Néel, L. (1932). Ann. de physique **17**, 5.

Neumann, F. E. (1885). *Vorlesungen über die Theorie Elasticität der festen Körper und des Lichtäthers* (Teubner, Leipzig, 1885).

Ofelt, G. S. (1962). J. Chem. Phys. **37**, 511.

Ovchinnikova, E. N., Oreshko, A. P. & Dmitrienko, V. E. (2025). *Physics-Uspekhi* **68**, 393-400.

Paolasini, L. (2014). de la Société Française de la Neutronique (SFN) **13**, 03002.

Paolasini, L., Vettier, C., de Bergevin, F., Yakhou, F., Mannix, D., Neubeck, W., Stunault, A., Altarelli, M., Fabrizio, M., Metcalf, P. A. & Honig, M. (1999). Phys. Rev. Lett. **82**, 4719.

Paolasini, L., Di Matteo, S., Vettier, C., de Bergevin, F., Sollier, A., Neubeck, W., Yakhou, F., Metcalf, P. A. & Honig, J. M. (2001). J. Elect. Spectr. and Rel. Phen. **120**, 1.

Pourovskii, L. V., Mosca, D. F., Celiberti, L., Khmelevski, S., Paramekanti, A. & Franchini, C. (2025). Nat. Rev. Materials. https://doi.org/10.1038/s41578-025-00824-z.

Scagnoli, V. & Lovesey, S. W. (2009). Phys. Rev. B **79**, 035111.

Scagnoli, V., Staub, U., Bodenthin, Y., de Souza, R. A., García-Fernández, M., Garganourakis, M., Boothroyd, A. T., Prabhakaran, D. & Lovesey, S. W. (2011). Sci. **332**, 696.

Shull, C. G. & Samuel Smart, J. (1949). Phys. Rev. **76**, 1256.

Tanaka, Y., Kojima, T., Takata, Y., Chainani, A., Lovesey, S. W., Knight, K. S., Takeuchi, T., Oura, M., Senba, Y., Ohashi, H. & Shin, S. (2010). Phys. Rev. B **81**, 144104.

Templeton D. H. & Templeton, L. K. (1985). Acta Crystallogr. A **41**, 133.

Templeton D. H. & Templeton, L. K. (1986). Acta CrystallogrA **42**, 478.

Winkler, R. & Zülicke, U. (2025). arXiv: 2405.20940.

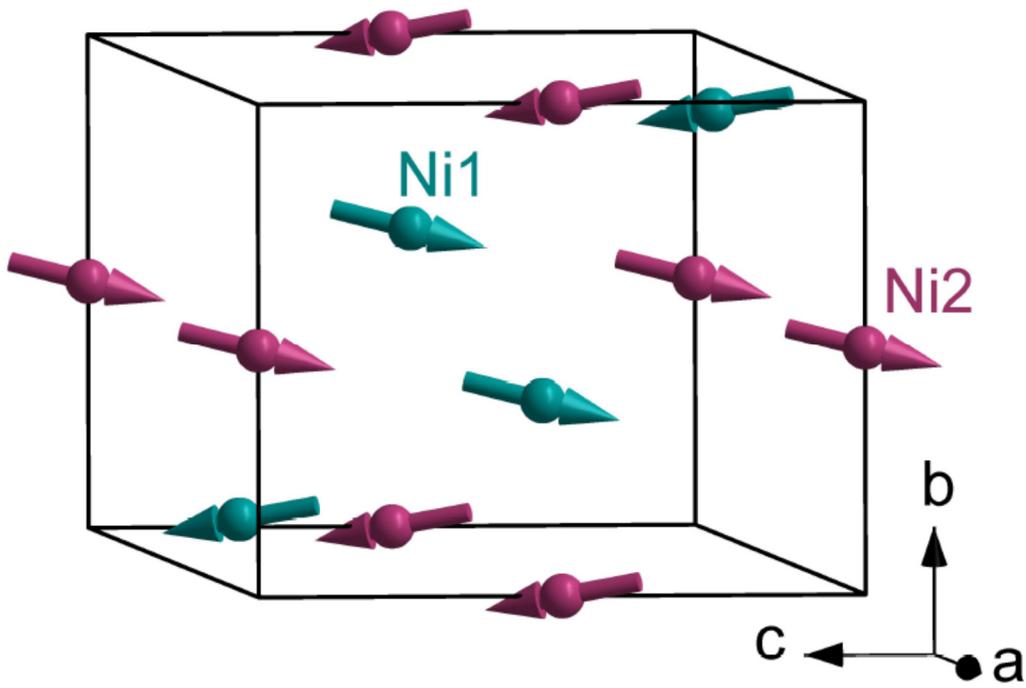

**FIG. 1**. Axial dipoles in $Li_2Ni_3P_4O_{14}$ using magnetic symmetry $P2_1/c$ (No. 14.75, propagation vector $\mathbf{k} = 0$, Bilbao). They are in Wyckoff general positions 4e for Ni1 (red arrows) and 2c for Ni2 (green). Chikara *et al*. (2025) constrained the two sites to have identical dipole moments and the inferred values are used in the figure.

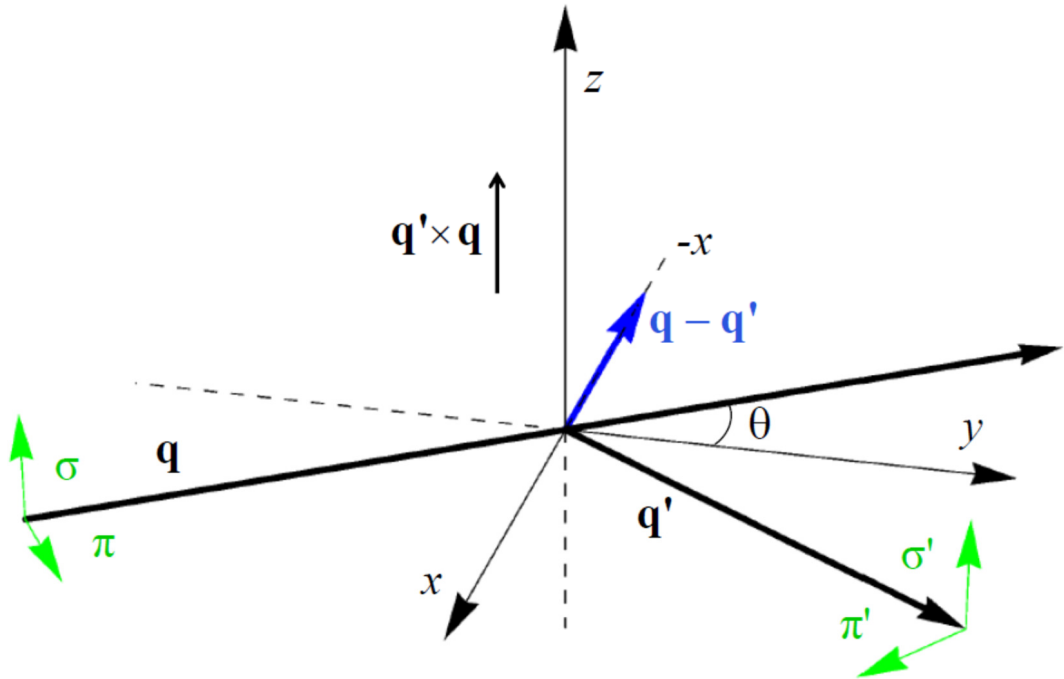

**FIG. 2**. Primary (σ, π) and secondary (σ', π') states of polarization. Corresponding wavevectors **q** and **q**′ subtend an angle 2θ. The Laue condition for diffraction is met when **q** − **q**′ coincides with a reflection vector ($h$, $k$, $l$) of the monoclinic reciprocal lattice. Crystal vectors that define local axes for Ni ions (ξ, η, ζ) and the depicted Cartesian (x, y, z) coincide in the nominal setting of the crystal.